\documentclass[12pt]{article}

\usepackage{graphicx}
\usepackage{epstopdf}
\usepackage[body={17.5cm, 21cm},right=2cm]{geometry}
\usepackage{amssymb}
\usepackage{amsmath}
\usepackage{cite}
\usepackage{hyperref}
\usepackage{subcaption}
\numberwithin{equation}{section}

\newcommand\eea{\end{eqnarray}}
\newcommand\bea{\begin{eqnarray}}

\def\beq{\begin{equation}}
\def\eeq{\end{equation}}

\newcommand{\be}{\begin{equation}}
\newcommand{\ee}{\end{equation}}
\newcommand{\ba}{\begin{align}}
\newcommand{\ea}{\end{align}}
\newcommand{\bg}{\begin{gather}}
\newcommand{\eg}{\end{gather}}
\newcommand{\bseq}{\begin{subequations}}
\newcommand{\eseq}{\end{subequations}}

\begin{document}

\vspace{5mm}
\vspace{0.5cm}
\begin{center}

\def\thefootnote{\fnsymbol{footnote}}

{\Large \bf Positive curvature and scalar field tunneling in the landscape}
\\[0.5cm]

{\small Bart Horn\footnote{Email: bhorn@physics.ucsd.edu}}

\vspace{.2cm}

{\small \textit{
High Energy Theory Group, Department of Physics,\\
University of California at San Diego, La Jolla, CA, 92093 USA}}
\end{center}

\vspace{.8cm}

\noindent We present a model of vacuum tunneling through a classically forbidden region where a scalar field changes its value simultaneously over the entire volume of a (meta)stable ancestor vacuum with spherical curvature.  The tunneling leaves the geometry unchanged but rearranges the energetic contributions of the matter sources, leading to an inflating solution with residual positive curvature.  We show that there exists a parametric regime where this solution is self-consistent and dominates the overall tunneling rate.  We conclude that an experimental detection of positive curvature, while not necessarily likely, therefore does not rule out the possibility that our present observer patch originated from semiclassical vacuum tunneling in a string or field theoretic landscape.

\noindent

\vspace{0.5cm}  \hrule
\def\thefootnote{\arabic{footnote}}
\setcounter{footnote}{0}




\section{Introduction}

Spatial curvature on cosmological scales, if present, is a potential indicator of pre-inflationary physics with important implications for constraining or supporting models of primordial cosmology.  While many studies of precision cosmological data have focused on models of, and parameters related to, inflation and the seeds of structure, recent and upcoming cosmological surveys also have the power to measure the large scale geometry and evolution of our observer patch at late times.  A detection of spatial curvature on large scales would provide important information about the duration and initial conditions of inflation, and if nonzero, it is perhaps one of the more significant single numbers that we can measure.  The experimental limits are expressed in terms of the present value of the parameter 
\begin{equation}
\Omega_K = -\frac{3k}{8\pi G_N a^2 H^2}\Bigg|_{0}
\end{equation}
where all quantities are measured at the present epoch (in which case the scale factor $a$ is usually normalized to the value $1$), $k$ is the inverse square of the radius of curvature, $G_{N}$ is Newton's constant, and $H$ is the Hubble scale.  This parameter $\Omega_{K}$ is constrained to be nearly flat at the $1-2\%$ level by the CMB temperature and polarization data alone, and to the level $5 \times 10^{-3}$ if the CMB data are combined with data from baryon acoustic oscillation surveys\cite{vanEngelen:2012va, Ade:2015xua}.  By convention, a negative value for $\Omega_K$ corresponds to positive curvature or spherical spatial slicing, and a positive value corresponds to negative curvature or hyperbolic/open slicing.  Since a value $\Omega_K \sim 10^{-5}$ is expected to be generated at the same time as the inflationary power spectrum, any observed value for this parameter must be larger than this in order to have any significance as a detection of initial conditions, or in order to have an unambiguous meaning as a gauge-invariant observable at all\cite{Vardanyan:2009ft,Bull:2013fga}.

While inflating models with either sign of $\Omega_{K}$ have been constructed in the literature (see e.g.\cite{Bousso:1998ed, Linde:2003hc}), top-down theory considerations (beginning with \cite{Freivogel:2005vv}) seem to favor a positive value for $\Omega_K$, i.e. negative (open or hyperbolic) spatial curvature, which can arise very simply as a consequence of semiclassical false vacuum tunneling.  This process was first studied in the presence of gravity by Coleman and De Luccia \cite{Coleman:1980aw}, and the probability of false vacuum tunneling of a scalar field through a classically forbidden region is related to the Euclidean action for a ``bounce'' solution where the scalar interpolates from the false to the true vacuum and back again.  For an ancestor vacuum that is itself maximally symmetric in four dimensions, the tunneling rate will be dominated by an $O(4)$ symmetric scalar bounce solution, and upon analytic continuation to Lorentzian space, if the nucleated bubble is much smaller than the curvature radius, then the interior will have negatively curved spatial slicing.  If the measure on the landscape favors inflation to last just as many e-foldings as are necessary to solve the horizon and flatness puzzles, this residual curvature may persist to the present day and even be measurable on the largest scales \cite{Freivogel:2005vv}.

More recently, it has been stated \cite{Freivogel:2005vv, Batra:2006rz, Kleban:2012ph, Guth:2012ww, Freivogel:2014hca} that a detection of positive curvature would falsify (or at least provide strong evidence against) the possibility that our current observer patch originated within a string or field theoretic landscape.  Our purpose in the current work is to demonstrate a counterexample to this statement.  While we do not disagree with the technical conclusions of these works, which predict $\Omega_K \geq 0$ starting from certain fairly generic models of the landscape, it is worth examining and clarifying the assumptions being made in order to look for interesting loopholes.  The first is the assumption that the ancestor vacuum needs to obey maximal symmetry in four dimensions.  The second issue is that in known examples it can be difficult to make the tunneling take place over an entire spherical region: in the context of transitions between De Sitter minima, in order to get positive curvature, this would correspond to a tunneling acting over an entire closed slice of De Sitter, which is larger than a single observer static patch.  

This case was discusssed in detail in \cite{Batra:2006rz} and it is worth reviewing briefly why De Sitter to De Sitter transitions do not constitute a good way to generate an inflating solution with positive curvature.  For the transition over the entire closed slice (known as the Hawking-Moss instanton \cite{Hawking:1981fz}) to dominate the decay rate, rather than bubbles much smaller than the De Sitter radius, the potential must be broad and flat with $V'' \lesssim H^2$ at the top of the potential.  In this case the universe will be in an eternally inflating initial condition immediately following the tunneling, and then quantum fluctuations will erase all memory of the semiclassical event.  Furthermore, the interpretation of the Hawking-Moss process as a tunneling amplitude is less than straightforward for several reasons -- the interpretation of the instanton as a thermal fluctuation to the top of the potential barrier no longer makes sense once it extends outside a single static patch, the amplitude depends only on the initial and final values of the potential and seems not to know about the path taken in field space, and the presence of additional negative eigenvalues in the second variation of the action around the semiclassical solution affects the sign of the Euclidean action and makes the final interpretation of the path integral as a tunneling rate unclear\cite{Weinberg:2006pc, Brown:2007sd, Masoumi:2012yy}.

In the current work we avoid these problems by having the semiclassical tunneling take place on a perturbatively stable spherical ancestor, which may be either in a static or a periodically oscillating regime.  Our starting point is the `simple harmonic universe' (SHU) recently investigated in \cite{Graham:2011nb, Graham:2014pca}.  While there our focus was on whether this class of solutions may be stable at the perturbative and non-perturbative levels, here our goal is to engineer a metastable model that decays into an inflating solution with positive spatial curvature.  Although we do not have a fully worked UV complete model deriving this solution or its decay from string or field theory, and it would be interesting to pursue this in more detail, it nevertheless serves as a useful toy model for exploring the possibility of generating residual positive curvature from vacuum transitions, and serves to investigate whether general relativity and/or quantum field theory imposes any general obstruction to this possibility.  

While this is not the first work to present an example of a tunneling process that leads to an expanding universe with positive curvature, we feel it is worth revisiting this topic for several reasons.  Many of the previous proposals for generating positive curvature depend upon a tunneling event that alters the geometry of the universe.  In \cite{Atkatz:1981tk} the scale factor $a$ tunnels through a barrier in an effective potential $V{eff}(a)$ from a static configuration into a hot Big Bang phase, although the origin of the matter sector supporting the necessary equation of state and whether it is consistent (e.g.\ with the null energy condition) are not discussed in detail.  In \cite{Hartle:1983ai, Kallinin} the scale factor tunnels from a dust- or radiation- dominated phase into an inflating one, and in \cite{Dabrowski:1995jt} the scale factor tunnel from a dust-dominated phase to an oscillating SHU solution.  In these cases, however, there is only a small probability that the universe will tunnel, and most of the time the universe will simply recollapse to a singularity.  In \cite{Vilenkin} a solution is presented where the scale factor of the universe `tunnels from nothing' into a positively curved inflating solution.  In all these cases the ancestral geometry (if any) is completely or partially erased in the process, or there is no semiclassical ancestor to speak of.  The model we will present here takes a different approach, where the tunneling event does not involve the spacetime geometry, but rather uses a scalar field tunneling to rearrange the energy of various matter sources.  We will find that this can be accomplished self-consistently without encountering singularities in the Euclidean solution or violating the null energy condition.  There is therefore no general reason for an observation of positive curvature to preclude the existence of a landscape.  Furthermore, since the simple harmonic ancestor is perturbatively stable, at this level it is a local attractor in the space of solutions.

Another point worth making, and that has been discussed already in the literature, is that if the landscape contains a regime of eternal inflation, it should be possible and even mandatory that some terminal vacua will have positive curvature:\  the authors of \cite{Brown:2011ry} argue that the entire landscape (outside a set of measure zero) can be populated quantum mechanically this way, and the authors of \cite{Buniy:2006ed} explore whether purely quantum transitions may dominate and even favor positive curvature in certain corners of the landscape.  In this regime, however, as for the Hawking-Moss transition considered and rejected in \cite{Batra:2006rz}, the memory of the ancestor vacuum will be wiped out completely by quantum fluctuations, and we cannot calculate decay rates without some prior assumptions on the measure.  For the model we consider here, on the other hand, the tunneling rate is related to a semiclassical Euclidean solution and therefore can be related to other parameters in the model in a straightforward way.  We should emphasize that we do not know what the initial conditions on the landscape are that would lead to a simple harmonic ancestor, or whether this ancestor vacuum is typical or popular in any sense, but rather if the initial conditions do lead somehow to this class of solutions, they can then be long lived enough for the semiclassical transition to a positively curved inflating solution to take place.   While we do not know whether a detection of positive curvature is likely in a measure theoretic sense, we wish only to show that were a detection of positive curvature to be made, it would not falsify the hypothesis that the current phase of our universe originated as a semiclassical tunneling event, nor would it rule out the possibility that our current vacuum is part of a more extensive landscape.

This paper is organized as follows: In Section \S 2 we review the simple harmonic universe solution and present our tunneling instanton, which mediates between the simple harmonic ancestor and an inflating universe with positive curvature.  In \S 3 we check that there is a parametric regime where our solution is self-consistent and dominates the total tunneling rate.  We will find that while it is consistent to neglect the effects of gravity at the level of the background evolution, it becomes necessary to consider the effects of gravitational backreaction when considering fluctuations around the semiclassical event.  We conclude in \S 4 and briefly indicate the outlook for experimental measurements of $\Omega_{K}$ and further directions for investigation.

\section{CDL bounce solution from SHU to De Sitter}

Our starting point is the simple harmonic universe (SHU) studied in \cite{Graham:2011nb, Graham:2014pca}, consisting of positive spatial curvature undercancelled by a frustrated string network, a negative cosmological constant, and a source of matter behaving like a frustrated network of domain walls, with $p/\rho = - 2/3$ at the homogeneous level.  At the level of linearized perturbations, all matter sources are assumed to have a positive sound speed squared.  The Friedmann equation for the background metric 
\beq
ds^2 = -dt^2 + a(t)^2 d\Omega_{3}^{2}
\eeq
is given by
\beq
\left(\frac{\dot{a}}{a}\right)^2 = -\frac{1}{a^2} + \frac{8 \pi G_N}{3}\frac{c_{str}}{a^2} + \frac{8 \pi G_N}{3}\frac{\rho_{0}}{a} + \frac{8\pi G_N \Lambda}{3} \,.
\eeq
Here we have set the curvature parameter $k$ to $1$, and $c_{str}$, $\rho_{0}$ set the energies of the matter sources with $p/\rho = -1/3, -2/3$ respectively.  The cosmological constant $\Lambda$ will be $<0$.  Using the equivalent parametrization
\beq
\left(\frac{\dot{a}}{a}\right)^2 = - \frac{K_{eff}}{a^2} + 2\sqrt{\frac{K_{eff}}{\gamma}}\frac{\omega}{a} - \omega^2  \, ,
\eeq
where $\omega = \sqrt{8 \pi G_N |\Lambda| /3}$, $K_{eff} = 1 - 8\pi G_N c_{str}/3$, and $\gamma = (3K_{eff}|\Lambda|)/(2\pi G_N \rho_{0}^{2})$, the solution is found to be
\beq
a(t) = \frac{\sqrt{K_{eff}/\gamma}}{\omega}\left(1 + \sqrt{1-\gamma}\cos \omega t \right)\,.
\eeq
Here $\gamma$ is constrained to be $0 < \gamma \leq 1$, with $\gamma \ll 1$ corresponding to a large hierarchy between the maximum and minimum values of the scale factor, and $\gamma \approx 1$ to the static, or nearly static, limit.  In \cite{Graham:2014pca} it was shown that this solution can be made to be stable at the level of linearized perturbations, for most values (in a measure theoretic sense) of the parameters when $K_{eff} \lesssim \gamma$\footnote{Note that in the presence of multiple matter sources, it is useful to orbifold the $S^{3}$ along the Hopf fiber to the Lens space $S^{3}/\mathbb{Z}_{k}$ in order to forbid certain classes of perturbations \cite{Graham:2014pca}.}.  It is not known whether it is possible to extend the proof of stability to the nonlinear or nonperturbative levels (see \cite{Mithani:2011en, Graham:2014pca, Mithani:2014toa} for further discussion on this topic); however, our goal in the present work is merely for the universe to last long enough for a particular tunneling process to take place.

The Euclidean action for the SHU is given by
\begin{equation}
\begin{split}
S_{E} &= \int d\tau \, 2\pi^2 a^3 \left(\frac{-1}{16\pi G_N}R + V_{tot}(\phi_i, a)\right)\\
&= \frac{3\pi}{4G_N} \int d\tau \, a\Big[a'^2 + \omega^2 a^2 -2\sqrt{\frac{K_{eff}}{\gamma}}\omega a + K_{eff}\Big]\\
&= \frac{3\pi}{4G_N}\frac{K_{eff}^{3/2}}{3 \gamma^{3/2}\omega^2}\left(2(3-2\gamma)\sqrt{\gamma} - 6(1-\gamma)\mbox{arcsech}\sqrt{1-\gamma}\right)
\end{split}
\end{equation}
Here the prime denotes the derivative with respect to Euclidean time (not conformal time), and in the last integral we have used in the integral the Euclidean solution
\begin{equation}\label{EuclideanSolution}
a(\tau) = \frac{\sqrt{K_{eff}/\gamma}}{\omega}\left(1 - \sqrt{1-\gamma}\cosh\omega \tau\right)\,,\qquad  |\tau| \leq \frac{1}{\omega}\mbox{arccosh}\left(\frac{1}{\sqrt{1-\gamma}}\right)\,.
\end{equation}
This becomes $S_{E} \approx \frac{\pi K_{eff}^{3/2}}{2 G_N \omega^2}$ for $\gamma \approx 1$, and $S_{E} \approx \frac{\pi K_{eff}^{3/2}\gamma}{5 G_N \omega^2}$ for $\gamma \ll 1$.  In \cite{Mithani:2011en} the Euclidean action was interpreted as a tunneling amplitude to zero scale factor.  While this instanton has a curvature singularity at the point $a \to 0$, it is certainly possible to regularize the singularity, for instance by adding a small amount of radiation, in which case the endpoint is a classically allowed region of small but finite scale factor, which then collapses.  Again, we stress that our goal at present is not to stabilize against nonperturbative decay, but rather to design a tunneling process in the SHU, and to make sure that this proceeds more quickly than any other possible metastability of the spacetime.  Regularizing the Euclidean solution also means that it is well-defined to include it in the path integral \cite{Vilenkin:1998pp}.

\begin{figure}
\begin{center}
\includegraphics[width = 3 in]{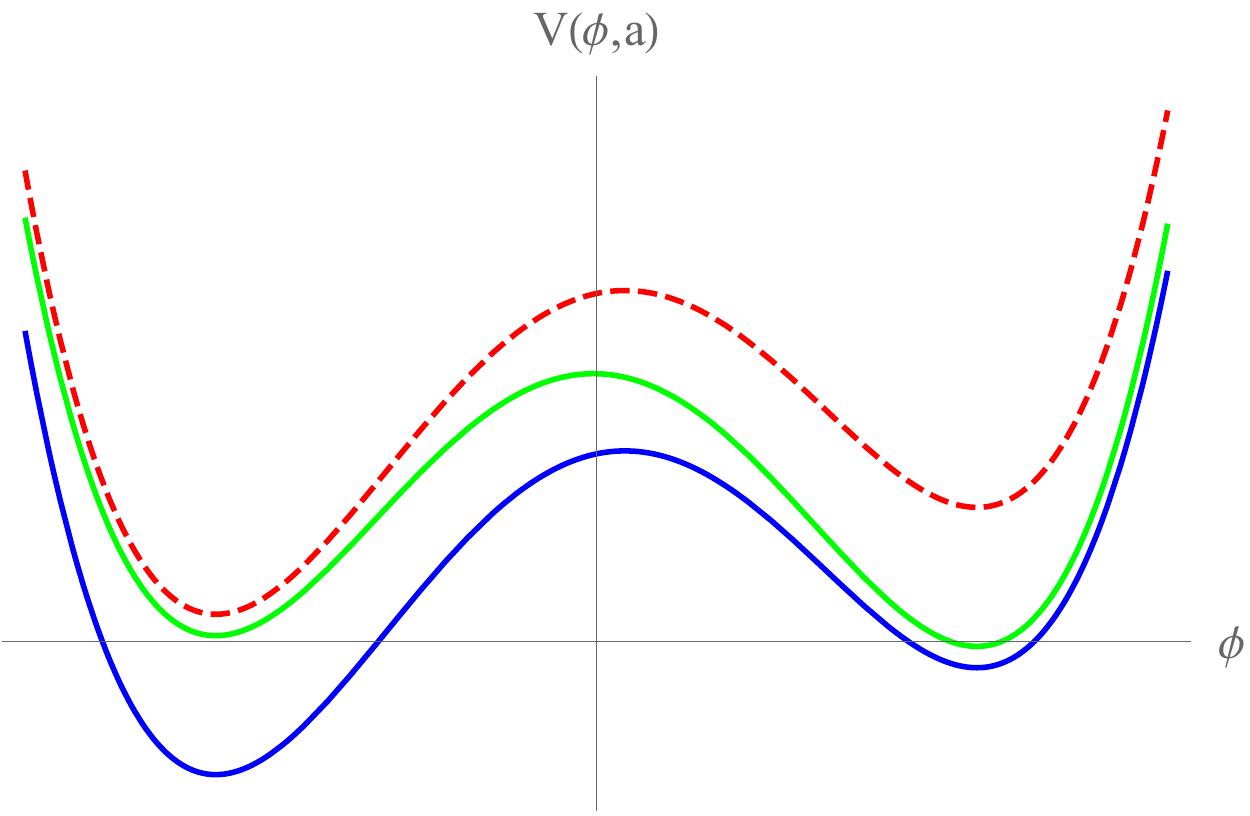}
\caption{The effective potential for different values of the scale factor.  $\phi_{i}$ is the local minimum on the left, and $\phi_{f}$ is on the right.  Green:\ $a = a_{min}$, Blue:\ $a = a_{max}$.  Dashed red:\ $a < a_{min}$ (classically forbidden)}
\label{fig:potential}
\end{center}
\end{figure}

Our strategy will be to engineer a tunneling solution from the SHU to an inflating spacetime, where the tunneling event happens over the entire spherical spatial slice of the ancestor, and in such a way that the equation of state of the matter is altered in the tunneling event, but not the geometry.  The universe will then enter a phase of inflationary expansion with residual positive curvature.  We can do this by requiring that both the vacuum energy and the energy of the matter source with $p/\rho = -2/3$ depend upon the vacuum expectation value of a single scalar degree of freedom, but not upon its kinetic energy, so that the Friedmann equation becomes
\beq
\left(\frac{\dot{a}}{a}\right)^2 = - \frac{K_{eff}}{a^2} + \frac{8 \pi G_N}{3}\left(\frac{\dot{\phi}^2}{2} + V_{tot}(\phi, a)\right) = - \frac{K_{eff}}{a^2} + \frac{8 \pi G_N}{3}\left( \frac{\dot{\phi}^2}{2}+\frac{V_1(\phi)}{a} + V_{2}(\phi)\right)  \, .
\eeq
We will suppose that the energetics are such that the transition is only energetically possible when $a \approx a_{min}$, as depicted in Fig.\ref{fig:potential}.  In addition, we will also add a radiation source whose energy depends on the vev of the same scalar.  This is necessary so that the transition will not be possible for $a < a_{min}$.  Although this region is classically forbidden for the Lorentzian solution, if it were not for the additional matter source the transition would be more likely to happen near the $a \to 0$ region of the Euclidean solution.  The Friedmann equation in Lorentzian signature will then become
\beq
\left(\frac{\dot{a}}{a}\right)^2 = - \frac{K_{eff}}{a^2} + \frac{8 \pi G_N}{3}\left(\frac{\dot{\phi}^2}{2} + V_{tot}(\phi, a)\right) = - \frac{K_{eff}}{a^2} + \frac{8 \pi G_N}{3} \left( \frac{\dot{\phi}^2}{2} + \frac{V_1(\phi)}{a} + V_{2}(\phi)+ \frac{V_{3}(\phi)}{a^4}\right)\, .
\eeq
The potentials $V_1(\phi), V_2(\phi), V_{3}(\phi)$ are assumed for simplicity to all have local minima at the same points $\phi_i$ and $\phi_f$ -- although this may be challenging to arrange from model-building point of view, there is no general obstruction to doing so, and we choose the values
\begin{equation}
\begin{split}
\phi &=\phi_i: \qquad \frac{8\pi G_N}{3}V_1 = 2\sqrt{\frac{K_{eff}}{\gamma}}\omega\,, \frac{8\pi G_N}{3}V_2 = -\omega^2\,, \frac{8\pi G_N}{3}V_3 \approx 0\,,\\
\phi &= \phi_f: \qquad \frac{8\pi G_N}{3}V_1 = 0\,, \frac{8\pi G_N}{3}V_2 = \frac{\frac{1}{2}+\sqrt{1-\gamma}}{1-\sqrt{1-\gamma}}\omega^2\,, \frac{8\pi G_N}{3}V_3 = \frac{1}{2}\frac{K_{eff}^{2}}{\gamma^{2}\omega^2}(1-\sqrt{1-\gamma})^3\,.
\end{split}
\end{equation}
The initial configuration is a static SHU with $a = a_{min} =  (1-\sqrt{1-\gamma})\sqrt{K_{eff}/\gamma}/\omega$, and the final configuration is an inflating solution starting with $a = a_{min}$.  The initial and final values are determined by requiring that $V_{tot}(\phi_{i,f}, a_{min}) = 0$ and $(\partial /\partial a)(V_{tot}(\phi_{i}, a) - V_{tot}(\phi_{f},a))|_{a=a_{min}} = 0$, so that the transition is possible only for $a = a_{min}$\footnote{Strictly speaking, it may be better to choose $V_{2}(\phi_{f}), V_{3}(\phi_{f})$ so that the final values are slightly smaller than the ones above, so that the tunneling is energetically possible for a range of values around $a_{min}$. We have reflected this in Fig.\ref{fig:potential}. This allows for the possibility of our initial conditions being slightly off, e.g. due to higher order effects from backreaction on the geometry or a mistuning of initial conditions.  It also allows for a region of slow-roll inflation to take place after the tunneling event has occurred.  We just need the tunneling to be most likely to happen at this point.}.  The inflating solution will have a small positive spatial curvature, and a radiation contribution which will dilute away to undetectable levels within a few e-folds.  The energy initially contained in the matter with $p/\rho = -2/3$ is used to raise the vacuum energy from positive to negative, and also to turn on the energetic contribution of the radiation source.  Note that we have also allowed for the possibility of a small but nonzero initial value for the radiation energy contribution, in order to regularize the singularity of the Euclidean solution.

We seek a solution to the Euclideanized equations of motion which interpolates between the initial and final minima.   The equations of motion are:
\begin{equation}
\begin{split}
\phi'' + 3\frac{a'}{a}\phi' &= \frac{V'_{1}(\phi)}{a} + V'_{2}(\phi) + \frac{V'_{3}(\phi)}{a^4}\\
\left(\frac{a'}{a}\right)^2 &= \frac{K_{eff}}{a^2} - \frac{8 \pi G_N}{3}\left(-\frac{\phi'^2}{2} + \frac{V_{1}(\phi)}{a} + V_{2}(\phi) + \frac{V_{3}(\phi)}{a^4}\right)
\end{split}
\end{equation}
Here, once again the prime denotes the derivative with respect to the Euclidean time $\tau$ (not with respect to conformal time).  Note that these are the same as the equations of motion for an $O(4)$ symmetric bounce in flat space, except that here, the radial coordinate is the Euclidean time $\tau$, and not the radial coordinate $s = \sqrt{\tau^2 + \vec{x}^2}$.

We can find a self-consistent solution with $a \approx a_{min}$, $a' \approx 0$ throughout the bounce, by solving the full equations perturbatively.  This corresponds to the regime where the Euclidean time scale for the scalar field evolution is much shorter than the Euclidean time scale for the scale factor to evolve appreciably.  Setting $a = a_{min}$ and $a' = 0$, the scalar equation becomes 
\beq\label{scalarOnlyEOM}
\phi'' = \frac{dV_{tot}(\phi, a_{min})}{d\phi} \, ,
\eeq
and the Hubble friction-like term can consistently be ignored as long as
\beq
\tau_{scalar} \sim \frac{\Delta \phi}{\phi'} \sim \frac{\Delta \phi}{\sqrt{V_{tot}}} \ll \frac{a}{a'}
\eeq
for the duration of the bounce.  Here $\Delta \phi$ is the scalar field excursion, and $V_{tot}$ is the total height of the potential.  We will check in the next section that there is a parametric regime where this condition can be self-consistently satisfied.  The Friedmann equation will hold for $a = a_{min}$ because
\beq\label{scalarHamiltonian}
\left(-\frac{\phi'^2}{2} + V_{tot}(\phi, a_{min})\right)
\eeq
is conserved given \eqref{scalarOnlyEOM}.  The gravitational backreaction is not small in this process, however, since the scalar equation and Hamiltonian constraint together imply the second Friedmann equation
\begin{equation}
\frac{a''}{a} = -\frac{8 \pi G_N}{3}\left(\phi'^2 + \frac{V_{1}(\phi)}{2a} + V_{2}(\phi) - \frac{V_{3}(\phi)}{a^4}\right)
\end{equation}
The right hand side will not vanish, and even though the Euclidean energy \eqref{scalarHamiltonian} for the matter sector is conserved, its equation of state varies dramatically during the bounce.  This is not unexpected: since the tunneling event involves the entire matter sector all at once, we should not expect the backreaction to be small.  Rather, the best we can do is to require that the scalar bounce is over before it has had time to affect the geometry appreciably.

The Euclidean action for the process can be estimated from the WKB formula
\begin{equation}\label{scalarBounceAction}
\begin{split}
S_E \sim \int d\tau \, a_{min}^3 \, V_{tot} &\sim \frac{\Delta \phi}{\sqrt{V_{tot}}} \, a^3_{min} \, V_{tot} \sim \Delta \phi \sqrt{V_{tot}} a^3_{min}\\ &\sim \Delta \phi \sqrt{V_{tot}} \left(\frac{\sqrt{K_{eff}\gamma}}{\omega}\right)^3 \sim \frac{\Delta \phi}{M_P}\frac{M^2_P}{\omega^2}K_{eff}^{3/2}\gamma\left(\frac{V_{tot}\gamma}{\omega^2 M_P^2}\right)^{1/2}\,.
\end{split}
\end{equation}
Here $V_{tot} \sim \omega^2 M_P^2/\gamma$ is the size of the other terms in the potential and therefore a natural estimate for the barrier height, but we leave the parametric dependence explicit for the sake of completeness.  We have written the parametric estimate in a way that is accurate both for $\gamma \to 0$ and for $\gamma \to 1$.  The contribution to the decay rate is then
\beq
\Gamma \propto e^{- \frac{\Delta \phi}{M_P}\frac{M^2_P}{\omega^2}K_{eff}^{3/2}\gamma \left(\frac{V_{tot}\gamma}{\omega^2 M_P^2}\right)^{1/2}}\,.
\eeq

We have considered the Euclidean solution in the approximation where only the scalar bounces and the background geometry is fixed.  This is valid near $\tau = 0$, when $a = a_{min}$.  At large $|\tau|$, however, the scale factor too will evolve, and the Euclidean solution will look like the solution in Eq.\eqref{EuclideanSolution}, up to small modifications due to the presence of the scalar bounce at $a = a_{min}$ and $\tau = 0$ having shifted initial conditions\footnote{Note that this is the minimum of the scale factor for the Lorentzian solution, and the maximum for the Euclidean solution.}.  The scalar bounce is therefore a multi-bounce solution, with the scalar bounce taking place over an entire timeslice of the much larger scale factor bounce, and the tunneling rate is derived by considering the difference in the Euclidean actions for the Euclidean simple harmonic universe with and without the scalar bounce.

Although the action for the scalar bounce can indeed be calculated by assuming that the background geometry is nearly fixed, since most of the volume of the Euclidean solution is dominated by the region where the scalar is fixed at $\phi = \phi_{i}$ and only the scale factor evolves, we might worry that a small shift $a \to a + \delta a$ in the initial conditions due to the scalar bounce might lead to a substantial contribution to $\delta S_{E}$ at large $|\tau|$.  To estimate this effect, we consider
\begin{equation}
S_{E} + \delta  S_{E} = 2\pi^2\int_{-\tau_{*} - \delta \tau_{*}}^{\tau_{*} + \delta \tau_{*}}d\tau \,\Big[a + \delta a\Big]\Big[(a + \delta a)'^2 + \omega^2 (a + \delta a)^2 - 2\sqrt{\frac{K_{eff}}{\gamma}}\omega (a + \delta a) + K_{eff}\Big] 
\end{equation}
Expanding this out order by order, the terms proportional to one power of $\delta a$ vanish everywhere except at the boundaries and when the scalar field is moving -- this latter part is just the scalar bounce contribution estimated in Eq.\eqref{scalarBounceAction}.  The contribution from the shift in the endpoints of the integral can be estimated as
\beq
\delta S_{E} \sim S_{E}\left(\frac{\delta \tau_{*}}{\tau_{*}}\right) \sim S_{E}\left(\frac{\delta a}{a}\right) \sim S_{E}\left(\frac{a''}{a}\right)(\delta \tau_{bounce})^{2} \sim S_{E}\frac{V_{tot}}{M_{P}^2}\frac{(\Delta \phi)^{2}}{V_{tot}} \sim S_{E}\frac{(\Delta \phi)^{2}}{M_{P}^{2}}
\eeq
Here $S_{E} \sim \frac{M_{P}^{2} K_{eff}^{3/2}\gamma}{\omega^2}$ is the size of the purely gravitational action, and so this correction can therefore be kept parametrically smaller than the bounce action as long as $\Delta \phi \ll M_P$.  Expanding to higher order, the terms proportional to $(\delta a / a)^2$ are proportional to $S_{E}(\Delta \phi / M_P)^4$, and this expansion continues to be under control.


\section{Consistency checks}

We now perform a set of consistency checks on our solution.  Specifically, we confirm that there is a parametric regime where the approximations made in finding the Euclidean solution above are self-consistent, and where the scalar bounce dominates the total decay rate of the spacetime.  We also analyze the second variation of the action around the classical solution -- here the effects of gravitational backreaction are not small; nevertheless, we argue that the Euclidean solution still can have a sensible interpretation as a vacuum transition amplitude. 

\subsection{Bounce dynamics and kinematics}

For self-consistency of our solution to the Euclidean equations of motion, we require that the Euclidean timescale $\tau_{bounce}$ for the scalar bounce be much faster than the timescale on which the scale factor changes.  We also require that it be much longer than the light-crossing time for the bubble.

The thickness of the bounce wall is given in terms of the field excursion $\Delta \phi$ and the barrier height $V_{tot}$ by
\begin{equation}
\tau_{wall} \sim \frac{\Delta \phi}{\sqrt{V_{tot}}}
\end{equation}
and unless the bounce is thin-walled (which is possible to engineer here but unnecessary to do so), this will also be the parametric timescale for the whole scalar bounce.  In order for the scalar bounce to proceed without any appreciable change in the geometry, we must therefore have
\begin{equation}
\tau_{bounce} \ll \frac{a_{min}}{K_{eff}^{1/2}} \to \frac{\Delta \phi}{M_P}\left(\frac{\omega^2 M_{P}^{2}}{\gamma V_{tot}}\right)^{1/2} \ll \epsilon \, ,
\end{equation}
where we have kept the extra parameter $\epsilon = \tau_{wall}/\tau_{bounce}$ for the sake of completeness.  Requiring that the bounce time be longer than the light crossing time means that
\begin{equation}\label{crossing}
a_{min} \sim \frac{\sqrt{K_{eff}\gamma}}{\omega} \ll \frac{\Delta \phi}{\sqrt{V_{tot}}} \to \frac{\Delta \phi}{M_P}\left(\frac{\omega^2 M_{P}^{2}}{\gamma V_{tot}}\right)^{1/2} \gg \sqrt{K_{eff}} \,.
\end{equation}
The bounce will be nonlocal if it occurs faster than the light-crossing time; however, it is unclear whether this is really a problem physically, since by the energy-time uncertainty relation it will still take much longer than the light crossing time to perform a measurement to determine that the tunneling has occurred\footnote{We thank E. Weinberg and A. Vilenkin for helpful discussions on this point.}.  This parametric limit is also necessary to forbid the nucleation of bubbles much smaller than the size of the spherical slice.  Suppose that a bubble of radius $R_{small} \ll a_{\min}$ (i.e. a standard Coleman De Luccia instanton) nucleates; the bubble does not see the curvature of the spatial slice and its radius will be set by balancing the wall tension with the energy difference $\epsilon V_{tot}$ between the false and true vacua.  Balancing these two contributions gives
\begin{equation}
R_{small} \sim \frac{\Delta \phi}{\epsilon \sqrt{V_{tot}}}\,.
\end{equation}
Here $\epsilon \ll 1$ corresponds to the limit of a thin-walled bubble, and since $\epsilon \lesssim 1$ the condition \eqref{crossing} guarantees that the bubble radius will be much larger than the size of the universe.  This shows that small bubbles therefore do not dominate the landscape, and a similar argument also shows that inhomogeneous tunneling events are energetically forbidden by gradient energy in this particular parametric limit.

To summarize this subsection, we have shown that in order for the scalar bounce to proceed over the entire volume of the sphere when $a = a_{min}$ and without any appreciable change in the scale factor, we must have
\begin{equation}
a_{min} \ll \tau_{bubble} \ll \frac{a_{min}}{K_{eff}} \to \sqrt{K_{eff}}\epsilon \ll \frac{\Delta \phi}{M_P}\left(\frac{\omega^2 M_{P}^{2}}{\gamma V_{tot}}\right)^{1/2} \ll \epsilon \lesssim 1\,.
\end{equation}
We can assume $\epsilon \sim 1$, $V_{tot} \sim \omega^2 M_{P}^{2}/\gamma$ for simplicity, in which case this simplifies to 
\begin{equation}
\sqrt{K_{eff}} \ll \frac{\Delta \phi}{M_{P}} \ll 1\,.
\end{equation}
These conditions can both be self-consistently satisfied as long as $K_{eff} \ll 1$.  Recall from the previous subsection that $\Delta \phi / M_P \ll 1$ is also already a necessary condition for the corrections to the Euclidean solution at large $|\tau|$ due to the presence of the scalar bounce to remain under control.

Note also that the universe will not be in an eternally inflating regime after the bounce, since
\begin{equation}
\frac{\frac{\partial^{2}}{\partial \phi^{2}}V_{tot}(\phi, a_{min})}{G_N V_{tot}} \sim \frac{M_{P}^{2}}{(\Delta \phi)^2} \gg 1\,,
\end{equation}
and so the memory of the semiclassical bounce is not erased by quantum fluctuations.

\subsection{Existence of negative eigenmodes around the semiclassical bounce}

We now analyze fluctuations around the bounce solution.  For nongravitational Coleman De Luccia tunneling of a scalar field $\phi$, in order for the Euclidean action to correspond to a decay rate, it is important that the second variational derivative of the action around the classical solution have exactly one negative eigenvalue\cite{Coleman:1987rm}.  For a general scalar field $\phi$ tunneling in a potential $V(\phi)$ in the absence of gravity, it can be proven that there is always exactly one such mode, and the single negative eigenvalue mode reflects the fact that the radius of the $O(4)$ symmetric bubble extremizes the Euclidean action.

To help build our intuition, consider a toy example where a scalar field in $0+1$ dimensions tunnels through a barrier in a potential $V(\phi)$ to a vacuum of lower energy.  The second variation of the action around the classical Euclidean solution is given by
\beq
\delta^2 S = \int d\tau \left\{ (\delta \phi) \left(-\frac{\partial^{2}}{\partial \tau^2} + V''(\phi_{cl}(\tau))\right)(\delta \phi)\right\}
\eeq
where $\tau$ is the radial coordinate, and $\phi_{cl}(\tau)$ is the classical bounce solution.  The problem therefore becomes equivalent to finding the number of bound states with negative energy in the Schr\"odinger ``potential'' given by $V''(\phi_{cl}(\tau))$.  These wavefunctions may be even or odd under time reversal.  By time translation invariance of the Lagrangian, there is a set of odd modes with zero energy proportional to $\phi'_{cl}(\tau)$, and there will be an even mode with lower (and therefore negative) energy.  If there exists another eigenmode with negative energy, depending on whether it is even or odd it can be combined with one of these two so as to create a path through the barrier which has smaller Euclidean action than the semiclassical solution, and therefore the Euclidean bounce does not have an interpretation as a sensible tunneling solution\cite{Coleman:1987rm}.  For a general scalar field $\phi$ tunneling in a potential $V(\phi)$ in the absence of gravity, it can be proven that there is always exactly one such negative eigenmode.  To help build intuition for this result, note that the potential $V''(\phi_{cl}(\tau))$ looks qualitatively like a pair of wells of width $\tau_{wall} \sim \frac{\Delta \phi}{\sqrt{V_{tot}}}$ and depth $1/\tau_{wall}^{2}$, where $\tau_{wall}$ is the thickness of the bubble.  The parametrics are therefore consistent with the presence of only a single bound state for a large class of potentials.  For nearly degenerate vacua the two wells are separated in space; while for a large energy difference, they may merge into a W-shape.  The single negative eigenvalue corresponds to the duration of the bounce extremizing the action.  Note that in the case of exactly degenerate vacua the negative eigenvalue goes to zero, and the tunneling is no longer interpreted as a decay rate.  The corresponding negative eigenmode becomes proportional to $|\phi'(\tau)|$, which is smooth at $\tau = 0$, in this limit.

\begin{figure}
\begin{centering}
  \begin{subfigure}[t]{0.48\textwidth}
    \includegraphics[width=\textwidth]{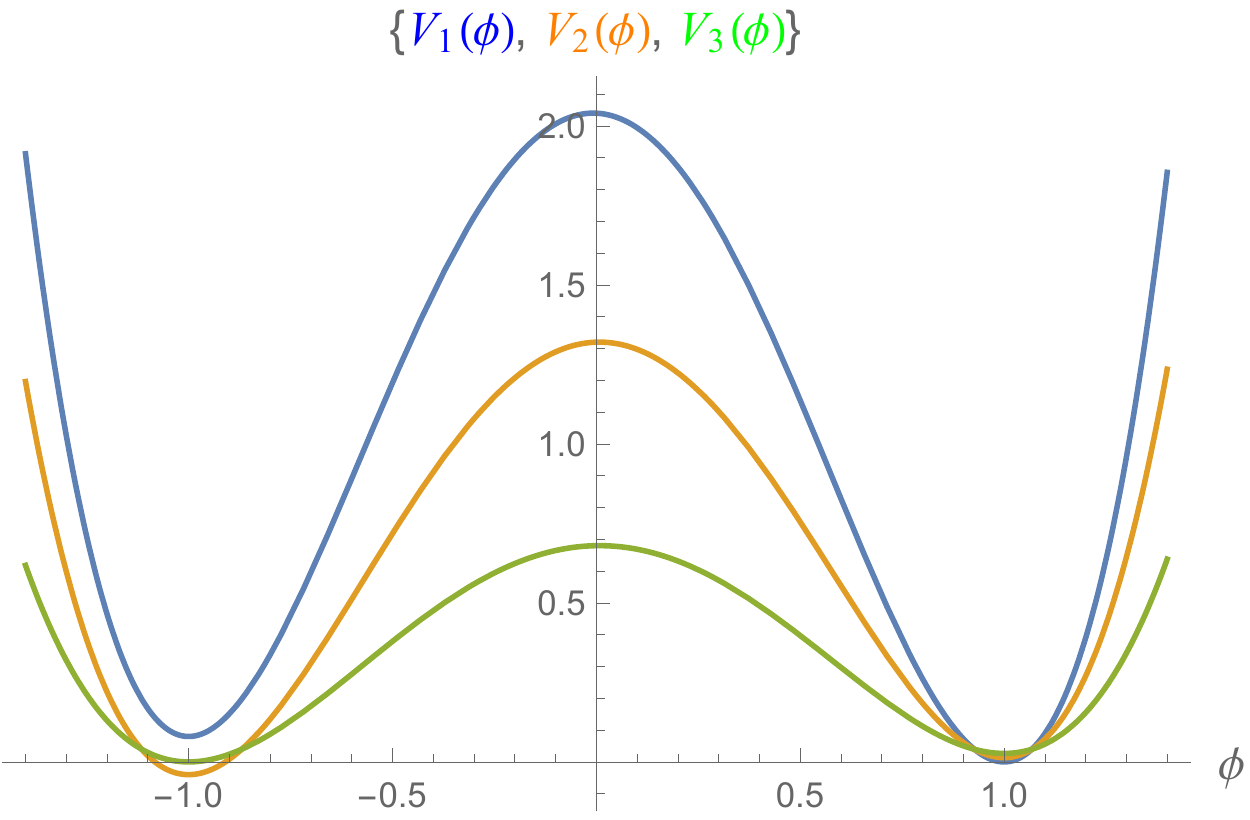}
    \caption{The potentials $V_{1}, V_{2}$ and $V_3$, with $a = 1$, are chosen so that the sums are equal before and after the bounce.  Note that $V_{1}$ downtunnels, and $V_{2}, V_{3}$ uptunnel.}
    \label{fig:1}
  \end{subfigure}
  \begin{subfigure}[t]{0.48\textwidth}
    \includegraphics[width=\textwidth]{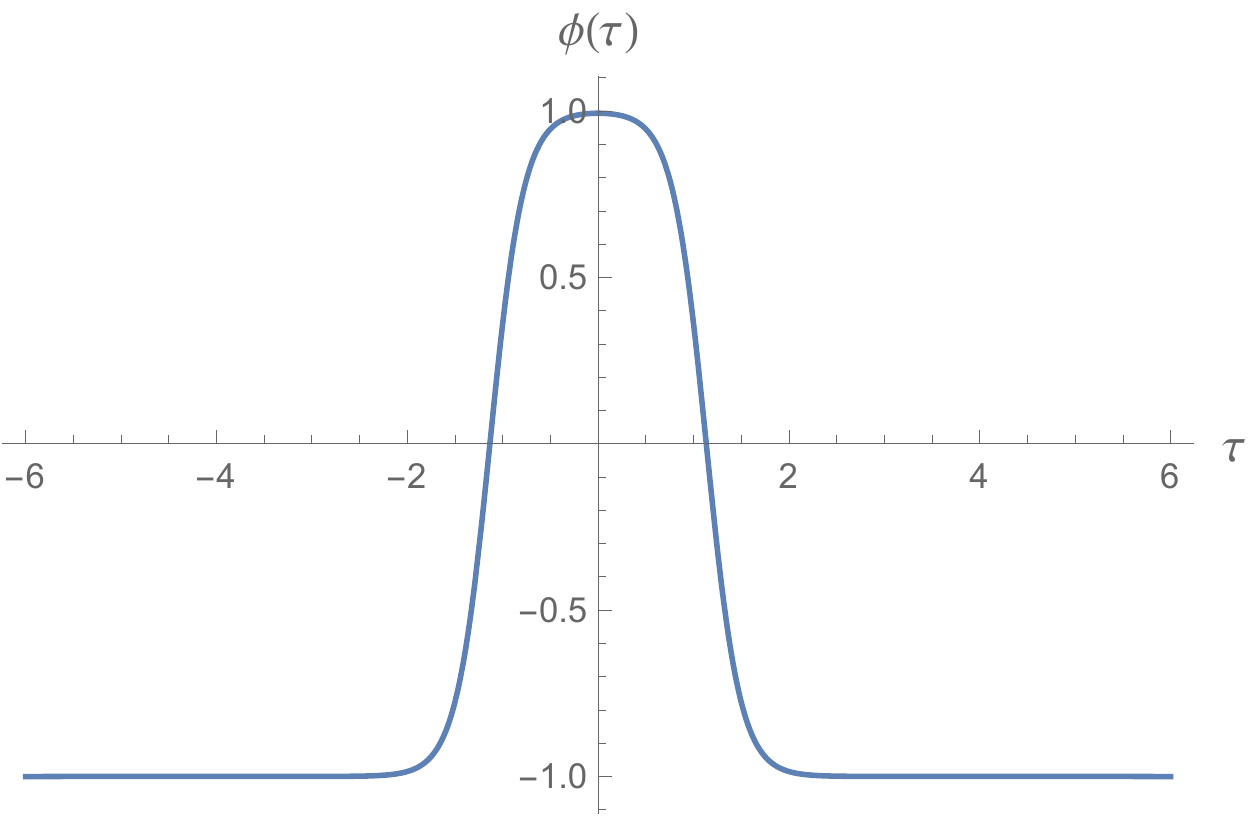}
    \caption{The corresponding field excursion in $\phi$ during the scalar field bounce.  The boundary conditions are such that $a'(0) = \phi'(0) = 0$, and $\phi(0)$ is chosen so that $\phi \to -1$ at large $|\tau|$.}
    \label{fig:2}
  \end{subfigure}
    \begin{subfigure}[t]{0.48\textwidth}
    \includegraphics[width=\textwidth]{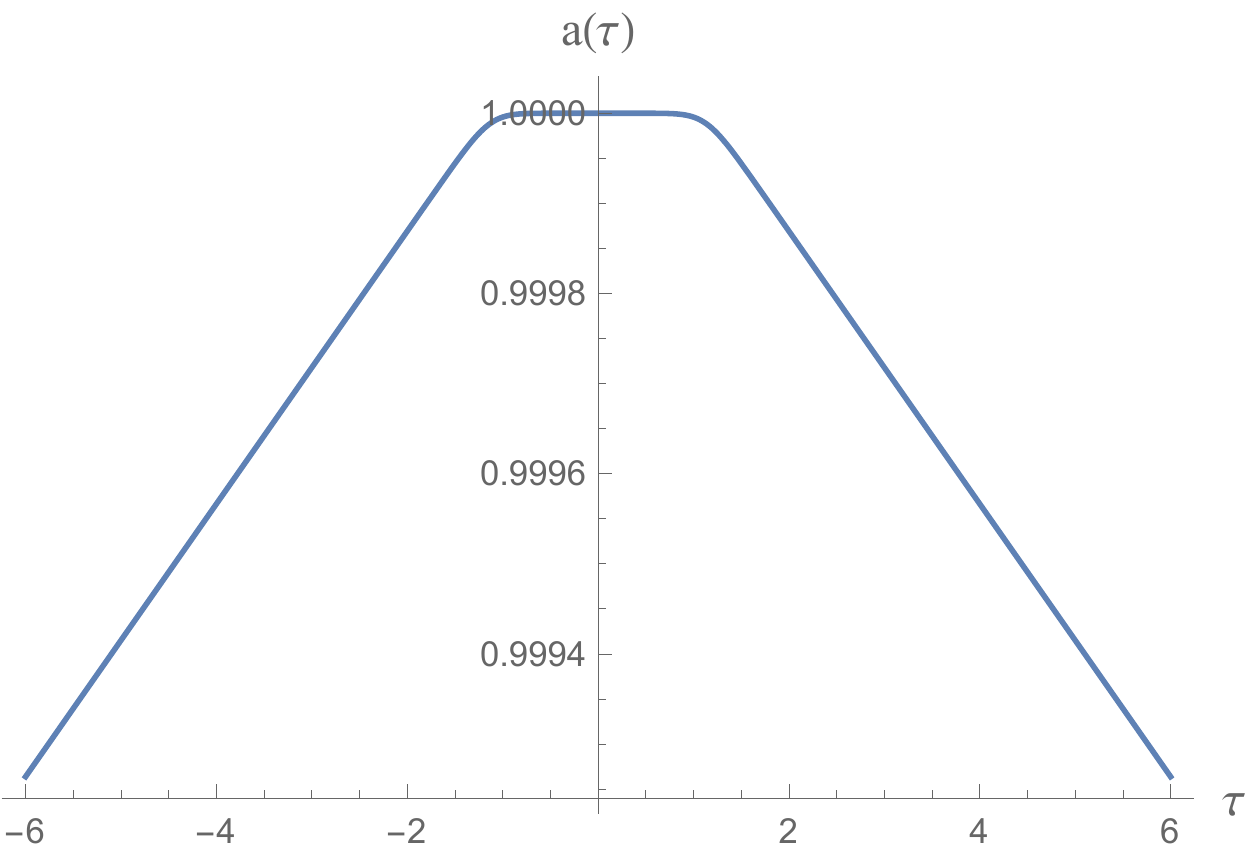}
    \caption{The scale factor is now allowed to evolve, with $\kappa = 8\pi G_N = 10^{-4}$.}
    \label{fig:3}
  \end{subfigure}
  \begin{subfigure}[t]{0.48\textwidth}
    \includegraphics[width=\textwidth]{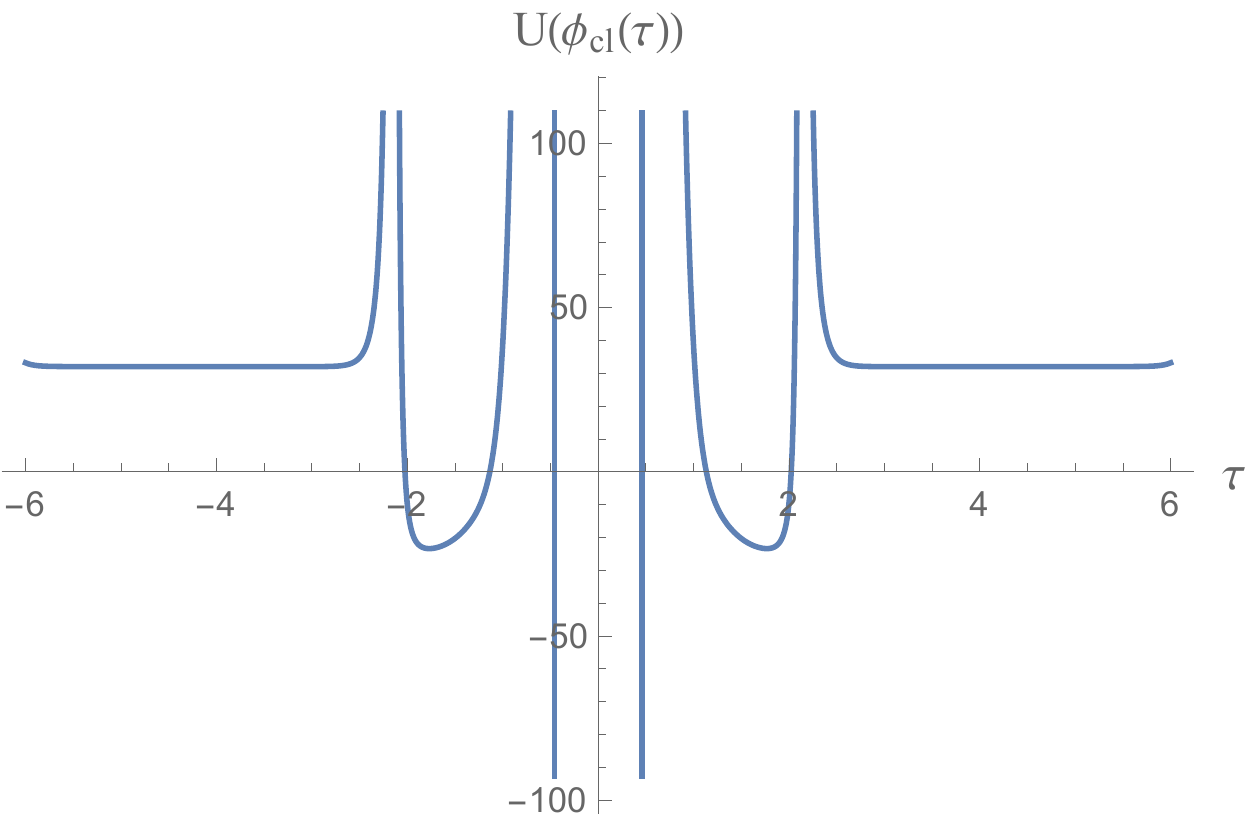}
    \caption{The effective potential for the quadratic fluctuations.  The localized minima support a single negative eigenvalue at $E = -12.218$.}
    \label{fig:4}
  \end{subfigure}
  \end{centering}
  \caption{The scalar bounce for a particular numerical example.  The potentials are given by $V_{1}(\phi) = 2(V(\phi)-V(1))$, $V_{2}(\phi) = \frac{4}{3}(V(-\phi)-V(-1))$, $V_{3}(\phi) = \frac{2}{3}(V(-\phi)-V(-1))$, where $V(\phi) = \phi^{4} +(0.001)\phi^{3} - 2\phi^2 - 3(0.001)\phi + 1$ has local minima at $\phi = \pm 1$.  We have suppressed the units, but parametrically the scales involved are $V_{tot}$ for $V_{1}, V_{2}, V_{3}$, $\Delta \phi$ for $\phi$, $\tau_{wall}\sim \Delta/\sqrt{V_{tot}}$ for $\tau$, and $V_{tot}/(\Delta \phi)^2$ for $U$.  Although the geometry does not change appreciably during the bounce, the potential for the fluctuations will receive nontrivial corrections when gravity is included.}
  \label{numericalbounce}
\end{figure}

When gravity is included, however, the derivation of the corresponding Schr\"odinger potential is more subtle due to the treatment of constraints in the off-shell action (see e.g. \cite{Lavrelashvili:1985vn, Tanaka:1992zw, Lavrelashvili:1998dt, Lavrelashvili:1999sr, Khvedelidze:2000cp, Gratton:2000fj, Lavrelashvili:2006cv, Dunne:2006bt, Lee:2014uza, Koehn:2015hga} for work on this problem).  We will apply the calculation of the effective Schr\"odinger potential for $O(4)$-symmetric perturbations in \cite{Lee:2014uza} to our current situation.  It is straightforward to include higher spherical harmonics as well, but these do not allow the appropriate bound states since the gradient energy is large when $V''/(G_{N}V_{tot}) \sim \frac{M_{P}^{2}}{(\Delta \phi)^2} \gg 1$\footnote{Also recall that the lowest few spherical harmonics may be orbifolded away}.  Writing our Euclidean action in the ADM formalism with the metric
\begin{equation}
ds^{2} = N d\tau^{2} + a^{2}d\Omega_{3}^{2}
\end{equation}
the Euclidean action is then
\begin{equation}
S_{E} = 2\pi^2 \int d\tau \sqrt{N}\Bigg\{a^3\Bigg[\frac{1}{2N}a'^2 + \frac{V_{1}(\phi)}{a} + V_{2}(\phi) + \frac{V_{3}(\phi)}{a^4}\Bigg] - \frac{3}{\kappa}\left(\frac{a a'^2}{N} + K_{eff}a\right)  \Bigg\}
\end{equation}
Here $N$ is the lapse, and $\kappa = 8\pi G_N$.  Expanding $N = 1+ 2A$, $a = a\sqrt{1+2\Psi}$, and $\phi = \phi + \Phi$, we can integrate out the non-dynamical variable $A$ and write the Lagrangian for the gauge-invariant combination $Y = \Phi - \frac{a}{a'}\phi'\Psi$:
\begin{equation}
\begin{split}
L_{E}(Y) &= \frac{a^3 a'^2}{2Q}\Bigg[Y'^2 + Y^2\left(\left(\frac{V_{1}''}{a} + V_{2}'' + \frac{V_{3}''}{a^4}\right) + \frac{\kappa a^2}{3Q}\left(\frac{V_{1}'}{a} + V_{2}' + \frac{V_{3}'}{a^4}\right)^2 \right. \\
& \left. + \frac{\kappa a \phi'}{3a' Q}\Bigg[K_{eff}\left(\frac{V_{1}'}{2a} + V_{2}' - \frac{V_{3}'}{a^4}\right) \right.\\
& \qquad \qquad \left. + \frac{\kappa a^2}{3}\left(\frac{V_{1}'V_2 - V_1 V'_2}{2a} - \frac{3}{2}\frac{V_{1}'V_{3} - V_{1}V_{3}'}{a^5} - 2\frac{V_{2}'V_{3}-V_{2}V_{3}'}{a^4}\right) \Bigg]\right)\Bigg]
\end{split}
\end{equation}
There will also be non-gauge invariant terms depending only on $\Psi$ that will turn out to be a total derivative.  Here $Q = a'^{2} - \frac{\kappa a^2}{6}\phi'^2$ can take on either sign, and we have made use of the background equations of motion.  Passing to the canonically normalized variable $\zeta = \frac{a^{3/2}a'}{\sqrt{Q}}Y$, the effective Schr\"odinger potential for $\zeta$ becomes:
\begin{equation}
\begin{split}
U(a_{cl}, \phi_{cl}) &= \frac{1}{2}\left(\left(\frac{V_{1}''}{a} + V_{2}'' + \frac{V_{3}''}{a^4}\right) + \frac{\kappa a^2}{3Q}\left(\frac{V_{1}'}{a} + V_{2}' + \frac{V_{3}'}{a^4}\right)^2 \right. \\
& \left. + \frac{\kappa a \phi'}{3a' Q}\Bigg[K_{eff}\left(\frac{V_{1}'}{2a} + V_{2}' - \frac{V_{3}'}{a^4}\right) \right.\\
&\qquad \left.+ \frac{\kappa a^2}{3}\left(\frac{V_{1}'V_2 - V_1 V'_2}{2a} - \frac{3}{2}\frac{V_{1}'V_{3} - V_{1}V_{3}'}{a^5} - 2\frac{V_{2}'V_{3}-V_{2}V_{3}'}{a^4}\right)\Bigg] \right. \\
&\left. +\left(\frac{3}{2}\frac{a'}{a} + \frac{a''}{a'}-\frac{Q'}{2Q}\right)' + \left(\frac{3}{2}\frac{a'}{a} + \frac{a''}{a'}-\frac{Q'}{2Q}\right)^2\right)
\end{split}
\end{equation}
Unfortunately, all the terms are parametrically the same size as the first, and so gravitational backreaction is not small for the fluctuations.  A specific numerical example which exhibits some generic features of this class of solutions is presented in Fig.\ref{numericalbounce}.  In particular we are interested in the behavior of $U(a_{cl}(\tau), \phi_{cl}(\tau))$, which is depicted in Fig.\ref{fig:4}.  The potential has a $-1/\tau^2$ singularity at $\tau = 0$ where $a'$, $\phi'$ vanish, though this region is too large negative to be visible in the figure.  There are a pair of asymptotia at $|\tau| \approx 0.45$ where $a'$ changes sign.  (Note that $a''(0)$ is positive because of the radiation term.)  There are then a pair of local minima, and a another set of asymptotia at $|\tau| \approx 2.13$ where $Q$ changes sign.  In terms of the parameters the localized minima have a duration $\tau_{wall}$ and a depth of $1/\tau_{wall}^{2}$, which is parametrically consistent with having a single negative mode with support in this region.  Solving numerically using the ``wag the dog'' method we find an eigenvalue with negative energy\footnote{Here by energy we mean the ``energy'' appearing in the Schr\"odinger equation, which in this case actually has dimensions of energy squared.}.  The corresponding mode can, however, be either even or odd, depending on how the two slowly varying parts of the wavefunction are stitched together at small $\tau$, and the singular region of the potential means that the wavefunction oscillates rapidly near $\tau = 0$ and may be consistent with either even or odd matching conditions without altering the value of the energy.  However, there is no longer a zero energy eigenmode with support in this region, now that gravity is included, and so the presence of a single odd eigenmode is not a problem.  


The pole at the midpoint of the bounce scales as $-1/\tau^{2}$, and therefore supports a continuum of eigenmodes taking on all negative values\footnote{There are also $1/\tau$ singularities at finite $\tau$, but these do not alter the story very much.}.  These oscillate infinitely many times near zero, and while wagging the dog by varying $E$ does not suffice to observe them, they can be observed for $E$ large negative by varying the precision of the cutoff in the potential near $\tau = 0$, in which case the tails at large $|\tau|$ switch between concave up and concave down as the number of oscillations near the singular point varies.  Consistent with the discussion in \cite{Lee:2014uza, Koehn:2015hga} we therefore find two classes of eigenmodes -- two slowly varying ones which are related to the one present in the flat space CDL bounce, of which one is odd and one is even, and an infinite set of rapidly varying ones that arise because the Euclidean action becomes unbounded below.  The rapid oscillations near $\tau = 0$ also mean that the single negative eigenvalue will broaden into a series of eigenmodes with nearly identical energies, but the differences between these nearly identical modes have support only in the rapidly oscillating regions.  These additional eigenmodes deserve to be understood further, but nevertheless we still expect that the Euclidean solution can correspond to a vacuum transition amplitude.  The additional eigenvalues are a purely gravitational effect and seem to be associated with the fact that our bounce involves a system of finite size.  While the sign of the bounce action may no longer correspond to a decay rate, the transition can be interpreted as a transition between degenerate minima.  See \cite{Lee:2014uza, Koehn:2015hga} for further discussions on this point.

We stress that we do not have an analytic proof that there is always only a single (even or odd) slowly varying negative eigenvalue, and it is possible that this fails in some corner of parameter space.  We seek only to demonstrate that it is possible for a fairly generic choice of parameters.  Note that had we not required the presence of the radiation contribution to forbid the transition at small $a$, we would also have a set of negative eigenvalues corresponding to the possibility that the tunneling event happens not at $a = a_{min}$ but in a more singular region of the Euclidean solution.  Again, this can be checked numerically, and in the examples we checked we have found a second set of slowly varying negative eigenvalues in this case.

As explained in the previous section, our scalar bounce solution takes place on a (possibly) metastable background, and so is therefore more precisely a multi-bounce solution; however, remember that it is the difference between the Euclidean actions with and without the scalar bounce that determines the scalar transition amplitude, and the only relevant negative eigenvalues are associated with the scalar bounce alone.  Note that we do not expect there to be any additional negative eigenvalues coming from the evolution of $a(\tau)$ alone at large $|\tau|$, since the Schr\"odinger potential $U(a_{cl}(\tau), \phi=\phi_{i})$ when $\phi$ remains at its initial value is positive everywhere thanks to the $V''$ terms in $U$.  Recall that the Euclidean solution without the scalar bounce was interpreted in \cite{Mithani:2011en} as a barrier penetration amplitude from the simple harmonic universe to a region of small $a$ (which is initially nonzero if we include the radiation contribution $V_{3}$).  Since there are no associated negative eigenvalues this does not correspond to a decay rate in the usual CDL sense; however, it can still be interpreted as a vacuum transition amplitude between degenerate solutions in a minisuperspace effective potential $V_{eff}(a)$\footnote{Strictly speaking, this is proportional to $a^4 V_{tot}(a, \phi_i)$ -- see \cite{Mithani:2011en}.}.

\subsection{Comparison with other decay modes}

The transition rate is proportional to
\beq
e^{-S_E} \sim e^{-\int d\tau a^3 V} \sim e^{-\frac{\Delta \phi}{M_P} \left(\frac{V_{tot}\gamma}{\omega^2 M_{P}^{2}}\right)^{1/2}\, \frac{M^2_P}{\omega^2}\, K_{eff}^{3/2}\gamma} \sim e^{-\frac{\Delta \phi}{M_P} \, \frac{M^2_P}{\omega^2}\, K_{eff}^{3/2}\gamma}\,,
\eeq 
which for $V_{tot} \sim \omega^2 M_{P}^{2}/\gamma$ and $\Delta \phi / M_P \ll 1$ means that the decay into an inflating solution proceeds exponentially more quickly than the purely gravitational process in \cite{Mithani:2011en}, which we take to be a typical action for any potential global gravitational instability of the entire SHU solution.


We might still worry about localized decay processes, such as those that might lead to the complete or partial collapse of the domain wall network, which itself presumably arises from some complicated metastable model in field theory.  Despite the model building challenges involved in constructing such a (meta)stable source, we expect that there is no model-independent reason why these cannot be overcome.  The exact tunneling rate will in general be model dependent; however, modeling the transition by a canonically normalized scalar field $\Phi$ that tunnels in a metastable potential $V_{\Phi}(\Phi)$, we can find a parametric solution where the $\phi$ decay proceeds more quickly.  A sufficient condition for this to occur is given by
\begin{equation}
\frac{\Delta \Phi \sqrt{V_{\Phi}}}{\Delta \phi \sqrt{V_{tot}}} \ll \frac{a^3_{min}}{\mbox{min}(a^{3}_{min}, R_{\Phi}^{3})}\,.
\end{equation}
where $\Delta \Phi$ is the field excursion between the false and true vacua, $V_{\Phi}$ is the potential barrier, and $R_{\Phi}$ is the radius of the bubble, if bubbles smaller than the size of the universe are allowed.  The right hand side is always $\gtrsim 1$, so even the weaker condition $\Delta \phi \sqrt{V_{tot}} \ll \Delta \Phi \sqrt{V_{\Phi}}$ will do.

Depending on the values chosen for the parameters, the lifetime of the universe can be tuned so that it lives for few, or many, bounces before tunneling.  Although the purpose of studying the SHU was originally to understand questions of stability over multiple bounces, here our purpose is merely to generate an inflating solution with positive curvature, and the universe does not even need to last as long as a single bounce (which may nevertheless be very long on particle physics timescales) before this happens.  In the case where the universe is to last for many bounces it is necessary that we choose values for $K_{eff} \lesssim \gamma$ so that the Universe is stable at the level of linearized perturbations.  Note that we have not been able to rule out the possibility of perturbative instabilities at the nonlinear level, although it may be possible to do so, particularly in the nearly static limit $\gamma \approx 1$ (see \cite{Graham:2014pca} for further discussion). However, the growth rate of such instabilities tends to be exponentially small in the size of the perturbations, and so it is still possible that even with perturbations and nonlinear couplings included, the ancestor vacuum is perturbatively an attractor and the total decay rate is dominated by the tunneling of $\phi$ discussed above.

\section{Conclusions}

In this paper we have demonstrated a proof-of-principle model where a semiclassical field-theoretic tunneling event mediates a transition from a stable spherically curved ancestor into an inflating solution with positive curvature.  The pre-inflationary geometry is left nearly intact in the tunneling and gives rise to the residual spatial curvature.  We do not know what the initial conditions are which would select this particular ancestor vacuum, nor do we claim that a detection of positive curvature is necessarily likely -- such questions await a more thorough understanding of probabilities on the landscape.  However, our model serves as an example that positive spatial curvature can be generated in a field theoretic landscape, and need not falsify the landscape scenario or erase all memory of the pre-inflationary vacuum.  Note also that the sub-Planckian field range $\Delta \phi$ was crucial for getting our model to work, suggesting that any subsequent stage of slow-roll inflation may also involve a small-field model as well, with constraints on model building for the inflationary sector.  From the theoretical standpoint, it would be interesting to explore these questions further, and also to find an explicit construction for the simple harmonic universe and this transition e.g. within string theory.

On the observational side, it would be interesting to embed more features of a realistic cosmology into this model, and to analyze the potential observational signatures.  In particular it would be interesting to perform a dedicated analysis of the effects of the positive curvature and the residual radiation source on precision observables -- for instance, the initial conditions from false vacuum tunneling may affect large-scale temperature anisotropies along the lines of \cite{Bousso:2013uia, Bousso:2014jca}.  Current bounds on the spatial curvature $\Omega_{K}$ are at the one percent level or below, and upcoming CMB experiments together with BAO data will sharpen this bound below the level of $10^{-3}$\cite{Abazajian:2016yjj, Finelli:2016cyd}.  Future 21cm surveys will also constrain the curvature\cite{Bull:2014rha}, and depending on assumptions for specific models of dark energy this bound may be sharpened even further, perhaps even closing the experimentally interesting window $|\Omega_K| \gtrsim 10^{-4}$ entirely\cite{Takada:2015mma, Leonard:2016evk}.  A detection of positive curvature, however unlikely, would provide evidence of physics from before the inflationary era, and on a more prosaic level may affect constraints on other precision parameters, such as the observed Hubble scale $H_0$.  Future experiments will explore this possibility and help us lift these degeneracies.

\section*{Acknowledgments}

The author would like to thank Peter Graham, Shamit Kachru, Surjeet Rajendran, and Gonzalo Torroba for helpful discussions and for collaboration on related issues leading to and inspiring this current work, and further to thank Adam Brown, Xi Dong, Raphael Flauger, Matt Kleban, Ali Masoumi, Sonia Paban, Alex Vilenkin, and Erick Weinberg for very helpful discussions.  This work was begun under the aegis of the Theory Group and the Institute for Strings, Cosmology and Astroparticle Physics at Columbia University, and subsequently the Weinberg Theory Group at the University of Texas at Austin.  This work is supported in part by the United States Department of Energy under DOE grants DE-FG02-92-ER40699 and DE-SC-0009919, by NASA under NASA ATP grant NNX10AN14G, and by the National Science Foundation under grants PHY-1512211586 and PHY-1620610.


\bibliographystyle{JHEP}
\renewcommand{\refname}{Bibliography}
\addcontentsline{toc}{section}{Bibliography}
\providecommand{\href}[2]{#2}\begingroup\raggedright

\end{document}